\documentstyle[12pt,epsf,a4,equations,mssymb]{article}

\def\mathrm{\rm}
\def\mathbf{\bf}

\newcommand{\be}{\begin{equation}}
\newcommand{\bea}{\begin{eqnarray}}
\newcommand{\beaN}{\begin{eqnarray*}}
\newcommand{\ee}{\end{equation}}
\newcommand{\eea}{\end{eqnarray}}
\newcommand{\eeaN}{\end{eqnarray*}}
\newenvironment{equation*}{\begin{displaymath}}{\end{displaymath}}
\newtheorem{Def}{Definition}
\newtheorem{Satz}{Satz}
\newtheorem{Theorem}[Satz]{Theorem}

\newtheorem{corollar}[Satz]{Corollary}
\newtheorem{Lemma}[Satz]{Lemma}

\newcommand{\const}{\mbox{const} }
\newcommand{\Scri}{\mbox{$\cal J$}}
\newcommand{\skizze}[1]%
  {\makebox[0.1pt]{}\\[0.5cm] {\it #1} \\[0.5cm] \makebox[0.1pt]{}}

\newlength{\Size}

\newcommand{\B}[1]{\bar{#1}}
\newcommand{\I}[1]{{}_{\mathrm{\mathbf{#1}\,}}{}}
\newcommand{\f}[1]{\underline{#1}}

\newcommand{\hG}{\mbox{$\hat{G}$}}
\newcommand{\hR}{\mbox{$\hat{R}$}}

\newcommand{\hT}{\mbox{$\hat{T}$}}
\newcommand{\Om}{\mbox{$\Omega$}}
\newcommand{\tC}{\mbox{$\tilde{C}$}}
\newcommand{\tG}{\mbox{$\tilde{G}$}}
\newcommand{\tT}{\mbox{$\tilde{T}$}}
\newcommand{\tg}{\mbox{$\tilde{g}$}}
\newcommand{\tn}{\mbox{$\tilde{\nabla}$}}
\newcommand{\tp}{\mbox{$\tilde{\phi}$}}
\newcommand{\tR}{\mbox{$\tilde{R}$}}

\newcommand{\thT}{\mbox{$\tilde{\hT}$}}

\newcommand{\N}[1]{ {\cal N}\I{#1} }

\floatsep\baselineskip

\begin{document}

\title{General Relativistic Scalar Field Models in the Large}

\author{{\bf Peter H\"ubner\footnotemark[1]} (pth@gravi.physik.uni-jena.de)\\
        Max-Planck-Gesellschaft Arbeitsgruppe Gravitationstheorie \\
        an der Universit\"at Jena \\
        Max-Wien-Platz 1 \\
        D-07743 Jena
        }
\newcommand{\oldthefootnote}{\thefootnote}
\renewcommand{\thefootnote}{\fnsymbol{footnote}}
\footnotetext[1]{This work is too a large extent part of my Ph.~D.\
      thesis which has been done at the Max-Planck-Institut f\"ur
      Astrophysik, Postfach 1523, D-85740 Garching}
\renewcommand{\thefootnote}{\arabic{footnote}}

\setcounter{page}{0}
\maketitle
\thispagestyle{empty}

\begin{abstract}
For a class of scalar fields including the massless Klein-Gordon field
the general relativistic hyperboloidal
initial value problems are equivalent in a certain sense. By using
this equivalence and conformal
techniques it is proven that the hyperboloidal initial value problem
for those scalar fields has
an unique solution which is weakly asymptotically flat. For data sufficiently
close to data for flat spacetime there exist a smooth
future null infinity and  a regular future timelike infinity.
\end{abstract}
\section{Introduction}
\subsection{General remarks}
A large open problem of classical general relativity is the
characterization of the structure of a spacetime by initial data. The
flat case, Minkowski spacetime, is geodesically complete. To the
other extreme the singularity theorems by R. Penrose and S. Hawking show
that the spacetime cannot be geodesically complete if the data are
large \cite{HaE73TL}.
\\
In the last years there has been remarkable progress in describing
what happens if one goes from data for flat space to large data: The
future of small data evolving in accordance with
the Einstein equation with various matter models as sources, vacuum
and Einstein-Maxwell-Yang-Mills, looks like the future of data for
flat space \cite{ChK93TG,Fr93as}. Nevertheless many problems are
still unsolved.
\\
Those results were significantly improved by D.~Christodoulou for
spherically symmetric models with a massless Klein-Gordon scalar field as
source. He was able to relate properties of the initial data to
properties of singularities. But even in this case of high symmetry
the questions left are still numerous as numerical
simulations by M.~Choptuik show \cite{Ch93ua}. He found very
interesting properties, the so called echoing effect, for models which
are in the parameter space of initial data near to the boundary which
separates regular from singular spacetimes.
\\
In this paper conformal techniques are used to analyze the
hyperboloidal initial value problem with scalar fields as
matter models --- for data near Minkowskian data the future of the
initial value
surface possesses a smooth future null infinity and a regular timelike
infinity, for large data a smooth future null infinity exists for at
least some time. In the second part of the introduction more about
conformal techniques and their application for a mathematical
description of asymptotically flat spacetimes will be said.
\\
Although the primarily treated matter model is that of the conformally
invariant scalar field, whose equations can be written as
\begin{subequations}
\label{model}
\begin{eqnarray}
\label{Wllngl}
  \tilde{\vphantom{\phi}\Box} \tp - \frac{\tR}{6} \, \tp & = & 0
\\
\label{EinstPhys}
  ( 1 - \frac{1}{4} \tp^2 ) \, \tR_{ab} & = &
       \left(
          (\tn_a \tp) (\tn_b \tp) - \frac{1}{2} \, \tp \, \tn_a \tn_b \tp
          - \frac{1}{4} \, \tg_{ab} (\tn^c \tp) (\tn_c \tp)
       \right),
\end{eqnarray}
the results obtained apply to a larger class of scalar field models,
given by the class of actions (\ref{ScalarAction}), including the massless
Klein-Gordon field, as shown in section \ref{SkalarAequiv}. Note that
an arbitrary factor can be absorbed into $\tp{}$ which changes the
coefficients in (\ref{EinstPhys}). My notational conventions are
described in the appendix, the $\tilde{\vphantom{H}}$ marks quantities
in the physical spacetime (see definition \ref{asymFlat}). The
energy-momentum tensor for the conformally invariant scalar field can
be written as
\begin{equation}
\label{TkonfS}
  \tT_{ab} = (\tn_a \tp) (\tn_b \tp) - \frac{1}{2} \, \tp \, \tn_a \tn_b \tp
        + \frac{1}{4} \, \tp^2 \tR_{ab} -
        \frac{1}{4} \, \tg_{ab}
          \left( (\tn^c \tp) (\tn_c \tp) + \frac{1}{6} \, \tp^2 \tR
          \right).
\end{equation}
\end{subequations}
The analytic investigation presented in this paper show the well
posedness of the initial value problem in unphysical spacetime, which
is a technical construct to ``compactified'' asymptotically flat
spacetimes in analogy to the compactification of the plane of complex
numbers ($\Bbb{R}^2$) into the Riemann sphere ($\Bbb{S}^2$).
\\
One goal
of this work was making myself familiar with the system in unphysical
spacetime as a preparation for numerical work showing
that the conformal techniques are well suited for a numerical
investigation of global spacetime structure and gravitational
radiation \cite{HuXXIP,Hu93nu}. To lower the computational
resources required these calculations have been done for spherical
symmetry. It is well known that spherically symmetric, uncharged vacuum
models are Schwarzschild. The inclusion of matter removes
that obstacle, the spacetime may evolve dynamically. Furthermore there
is no gravitational radiation in spherically symmetric models. Therefore
the matter model should also be a model for radiation. Scalar fields
with wave equations are choices for matter which model also radiation.
The conformally invariant scalar field has been
chosen since the matter equations are form invariant under rescalings
of the metric and an appropriate transformation of the scalar field
as the name already suggests.
\\
The scalar fields are interesting from the analytic viewpoint since
for the first time conformal techniques could be used for matter
models whose energy-momentum tensor has non-vanishing trace.
\subsection{Asymptotically flat spacetimes}
In this paper a geometrical, coordinate independent definition of
asymptotical flatness along the lines suggested by R.~Penrose
will be used. A more thorough discussion
of the ideas and the interpretation can be found at various places in
the literature, e.g.\ \cite{Ge76as,Pe64ct}. The
definitions of asymptotical flatness given in the literature differ
slightly. The following will be used here:
\begin{Def}
\label{asymFlat}
  A spacetime $(\tilde M, \tg_{ab})$ is called {\bf asymptotically
    flat} if there is another ``unphysical'' spacetime $(M,g_{ab})$
  with boundary \Scri{} and a smooth embedding by which $\tilde M$
  can be identified with $M-\Scri{}$ such that:
  \begin{enumerate}
  \item There is a smooth function $\Omega$ on $M$ with
    \begin{equation*}
      \Omega \mid_{\tilde M} > 0 \qquad \mbox{and} \qquad
      g_{ab} \mid_{\tilde M} = \Omega^2 \tg_{ab}.
    \end{equation*}
  \item On \Scri{}
    \begin{equation*}
      \Omega = 0 \qquad \mbox{and} \qquad \nabla_a \Omega \ne 0.
    \end{equation*}
  \item \label{nullCompleteness} Each null geodesic in $(\tilde
    M,\tilde g_{ab})$ acquires a past and a future endpoint on \Scri{}.
  \end{enumerate}
\end{Def}
Because of item~\ref{nullCompleteness} null geodesically incomplete
spacetimes like Schwarzschild are not asymptotically flat. The next
definition includes those spacetimes which have only an
asymptotically flat part:
\begin{Def}
\label{weakasymFlat}
  A spacetime is called {\bf weakly asymptotically flat} if
  definition~\ref{asymFlat} with the exception
  of item~\ref{nullCompleteness} is fulfilled.
\end{Def}
Definition~\ref{asymFlat} and~\ref{weakasymFlat} classify spacetimes,
they do not require that Einstein's equation is fulfilled. One would
like to know:
\\
Are they compatible with the Einstein equation with sources?
Neither definition~\ref{asymFlat} nor \ref{weakasymFlat} is in an
initial value problem form: A given spacetime is or is not classified
as asymptotically flat. But for a physical problem one would like to
give ``asymptotically flat data'' and have guaranteed that they evolve
into an at least weakly asymptotically flat spacetime.
\\
Nevertheless the geometrically description was extremely helpful in
analyzing asymptotically flat spacetimes and it can be successfully
used as guideline to construct a formalism which is better suited for
analyzing initial value problems. This method has been developed and
applied to various matter sources by H.~Friedrich~
\cite{Fr81on,Fr83cp,Fr85ot,Fr86op,Fr86ot,Fr88os,Fr91ot}. In this paper
it will be applied to general relativistic scalar field models.
\\
The idea is to choose a spacelike initial value surface in the
unphysical spacetime $(M,g_{ab})$ and to evolve it. The problems to
be faced are:
\\
For Minkowski space the unphysical spacetime $(M,g_{ab})$ can be
smoothly extended with three points, future $(i^+)$ and past  $(i^-)$
timelike infinity, the end respectively the starting point of all
timelike geodesics of $(\tilde M,\tg_{ab})$, and spacelike infinity
$(i^0)$,
the
end point of all spacelike geodesics of $(\tilde M,\tg_{ab})$. The point $i^0$
divides \Scri{} into two disjunct parts, future $(\Scri^+)$ and past
$(\Scri^-)$ null infinity. It is well known and has been discussed
elsewhere that there are unsolved problems in smoothly
extending a ``normal'' Cauchy hypersurface of $\tilde M$ to $i^0$ if the
spacetime has non-vanishing ADM mass. Certain curvature quantities
blow up at $i^0$, reflecting the non-invariance of the mass under
rescalings.
\\
By choosing a spacelike (with respect to $g_{ab}$) hypersurface $S$
not intersecting $i^0$ but $\Scri^+$ ($\Scri^-$) we avoid the problems
with $i^0$. $S$ is called a hyperboloidal hypersurface --- the
corresponding initial value problem is called a hyperboloidal initial
value (a detailed definition for the scalar field models is given in
section~\ref{HypInitValProblSec}, definition~\ref{HypInitValProbl}).
The domain of dependence $D(S)$ of
$S$ will not contain the whole spacetime. The interior of $S$
corresponds to an everywhere spacelike hypersurface in the physical
spacetime which approaches a null hypersurface $N$ asymptotically. If $N$
is a light cone $L$ then the domain of dependence of $S$ is $L$. Therefore
the hyperboloidal initial value problem is well suited to describe the
future (past) of data on the spacelike hypersurface $S$, e.~g.~a
stellar object and the gravitational radiation caused by its
time evolution. It is not well suited to investigate the structure
near $i^0$.
\\
But even for the hyperboloidal initial value problem there are
``regularity'' problems at \Scri{}: Transforming the Einstein equation
from physical to unphysical spacetime an equation ``singular'' for
$\Omega=0$ results. That problem is solved in this paper in analogy to
H.~Friedrich's work. A new set of equations
for the unphysical spacetime will be derived, its equivalence to the
Einstein equation on $\tilde M$ proven. This new set of equations is
used to prove the consistency of the hyperboloidal initial value
problem for scalar fields with (weakly) asymptotical flatness and the
existence of a regular future (past) timelike infinity
for data sufficiently close to data for Minkowski spacetime.
\section{Regularizing the unphysical field equations}
A first attempt for equations determining $(M,g_{ab})$ is the rescaled
form of the field equation in physical spacetime. A closer look at
the transformation of the Einstein tensor under rescalings
$g_{ab}=\Omega^2\,\tg_{ab}$,
\begin{equation}
\label{GabTransf}
  \tG_{ab} =
  G_{ab}
  + 2 \, \Omega^{-1} \left(
    \nabla_a \nabla_b \Omega - (\nabla^c \nabla_c \Omega ) \, g_{ab}
    \right)
  + 3 \, \Omega^{-2} ( \nabla^c \Omega ) ( \nabla_c \Omega ) \, g_{ab},
\end{equation}
shows that this first
attempt fails. Either there are terms proportional to $\Omega^{-2}$
and $\Omega^{-1}$, which need special care on the set ${\cal I}$ of
points where $\Omega=0$, including
\Scri{}, which is part of $M$. Or alternatively, the highest (second
order) derivatives of the metric, hidden in the Einstein tensor, are
multiplied by a factor of $\Omega^2$ and then the principal part of
the second order equation for the metric components vanishes on
${\cal I}$. This behaviour of an equation will be called singular on
${\cal I}$.
\\
In this section a system of equations without the singularity on
${\cal I}$ will be derived from the rescaled Einstein
equation by introducing new variables and equations.
\\
The set of equations together with the equations for the matter
variables may be a system with a very complicated principal part ---
as it is the case for a conformally invariant scalar field as
matter model. A procedure is carried out to simplify the principal
part to a form in which no equation contains both derivatives of
matter variables as well as derivatives of geometry variables and the
principal part of the subsystem for the geometry variables is the same
as for the vacuum case (``standard form''). All variables already
present in the vacuum case are called geometry variables.
\\
It is shown that the procedure works for the conformally invariant
scalar field. The procedure described does not use very restrictive
assumptions --- it is very general --- and may work for most matter
models, for which the unphysical matter equations can be regularized on
${\cal I}$.
\subsection{The geometry part of the system}
According to the definition of asymptotical flatness
(definition \ref{asymFlat}) the unphysical spacetime is connected with the
physical spacetime through the rescaling
\begin{equation}
  g_{ab} \mid_{\tilde M} = \Omega^2 \, \tg_{ab}.
\end{equation}
This rule also determines the transformation of the connection
and the curvature.
\\
Additionally the transformation of the matter variables $\tilde
\Phi{}$ under rescaling must be specified,
\begin{equation}
  \Phi \mid_{\tilde M} = \Phi[\tg_{ab},\Omega,\tilde \Phi].
\end{equation}
It is assumed that $\Phi{}$ has a smooth limit on \Scri{}, the
rescaled equations for the matter variables are regular on ${\cal
  I} $\footnote{In the general case it is not known how to achieve
  that.},
and there exists a tensor $T_{ab}$, which is independent of derivatives
of $\Omega{}$ and derivatives of curvature terms, fulfills
\begin{equation}
  T_{ab} \mid_{\tilde M} = \Omega^{-2}\, \tT_{ab},
\end{equation}
and has a limit on \Scri{} \footnote{From the definition of
  asymptotical flatness and the Einstein equation it follows that
  $\Omega^{-1}\, \tT_{ab}$ has a limit on \Scri{}~\cite{AsS80ni}. The
  faster fall off and the requirements on the form of $T_{ab}$ have
  technical reasons.}.
The conditions required may seem very restrictive but they
can be fulfilled for Yang-Mills fields~\cite{Fr91ot} and for the
conformally invariant scalar field.
\\[0.1cm]
The Riemann tensor will be split into its irreducible parts, the
conformal Weyl tensor
\begin{equation}
\label{ddef}
C_{abc}{}^d=:\Omega\,d_{abc}{}^d,
\end{equation}
the trace free part $\hR_{ab}$ of the Ricci tensor $R_{ab}$ and the
Ricci scalar $R$:
\begin{equation}
\label{Ralg}
  R_{abcd} =
  \Omega \, d_{abcd}
  + g_{c[a} \hR_{b]d} - g_{d[a} \hR_{b]c}
  + \frac{1}{6} g_{c[a} g_{b]d} R,
\end{equation}
A $\hat{\phantom{H}}$ is used as an indication for trace free parts of

tensors.
\\
The irreducible decomposition of the energy-momentum tensor is
\begin{equation}
\label{IrredT}
  T_{ab} = \hT_{ab} + \frac{1}{4} \, g_{ab} \, T.
\end{equation}
The irreducible parts transform under rescalings according to
$$
  \hT_{ab} = \Omega^{-2} \thT_{ab}
$$
and
$$
  T = \Omega^{-4} \tT.
$$
The vanishing of the divergence of $\tT_{ab}$ becomes
\begin{equation}
\label{transdivT}
  0 = \tilde\nabla^a \tilde T_{ab} =
  \Omega^4 \, \nabla^a \hT_{ab} +
  \frac{1}{4} \, \Omega^4 \, \nabla_b T +
  \Omega^3 \, T \, \nabla_b \Omega.
\end{equation}
For energy-momentum tensors with non-vanishing trace equation
(\ref{transdivT}) as an equation for the components of the
irreducible parts of the energy-momentum tensor $T_{ab}$
is singular on ${\cal I}$. Since (\ref{transdivT}) should be in same way
part of the matter equations problems in regularizing the matter
equations are to be expected.
\subsubsection{A regular system}
The part of (\ref{GabTransf}) proportional to $\Om^{-2}$ is a pure
trace, thus the $\Om^{-2}$ singularity is absent in the trace free equation.
A decomposition into the trace and the trace free part moves the
worst term into one equation.
\\
{}From the rescaling rule for the Ricci scalar and tensor,
\begin{equation}
\label{Rtrans}
  \tR = \Omega^2 R + 6 \, \Omega \, \nabla^a \nabla_a \Omega -
        12 \, (\nabla^a \Omega) \, (\nabla_a \Omega),
\end{equation}
and
\begin{eqnarray}
  \tilde{\hR}_{ab} & := & \tR_{ab} - \frac{1}{4} \tg_{ab} \tR
\nonumber \\
   & = & \hR_{ab} + 2 \, \Omega^{-1} \nabla_a \nabla_b \Omega
         - \frac{1}{2} \Omega^{-1} (\nabla^c \nabla_c \Omega) \,
           g_{ab},
\end{eqnarray}
$\tG_{ab} = \tT_{ab}$, ${\tG=-\tR}$, and ${\tT = \tG}$
it follows
\begin{equation}
\label{SpGl}
  \Omega \, R + 6 \, \nabla^a \nabla_a \Omega
    - 12 \, \Omega^{-1} \, (\nabla^a \Omega) \, (\nabla_a \Omega)
  = - \, \Omega^3 \, T
\end{equation}
and
\begin{equation}
\label{sfGl}
  \Omega \, \hR_{ab} + 2 \, \nabla_a \nabla_b \Omega -
  \frac{1}{2} (\nabla^c \nabla_c \Omega) \, g_{ab} =
    \Omega^3 \, T_{ab}.
\end{equation}
Equation (\ref{SpGl}) can be dealt with by the following lemma:
\begin{Lemma}
  From $\tR+\tT=0$ ($\hat{=}$ (\ref{SpGl})) at one point,
  $\hat{\tG}_{ab}=\hat{\tT}_{ab}$ ($\hat{=}$ (\ref{sfGl})), and $\tn^b
  {\tT_{ab}}=0$ $\tR+\tT=0$ follows everywhere.
\end{Lemma}
Proof:
\begin{equation*}
  \tn^a \tT_{ab} = \tn^a \thT_{ab} + \frac{1}{4} \tn_b \tT = 0.
\end{equation*}
Combined with
\begin{eqnarray*}
  0 & = & \tn^a \tG_{ab} \\
    & = & \tn^a \tilde{\hG}_{ab} + \frac{1}{4} \tn_b \tG
\end{eqnarray*}
gives
\begin{equation*}
  \tn_b (\tT + \tR) = 0,
\end{equation*}
i.e. $\tT + \tR $ is constant.
\\
Equation (\ref{SpGl}) will not be used any longer since $\tn^b
{\tT_{ab}}=0$ can be derived from the remaining equations, contract
(\ref{quaSysd}) or see the discussion following (\ref{IntegrBed}).
\\
In the following the Ricci scalar $R$ will be regarded as an
arbitrary, given function.
It fixes part of the gauge freedom on the transition from the
physical to the unphysical spacetime as follows:
The equations (\ref{SpGl}) and (\ref{sfGl}) are invariant under
rescalings ${(g_{ab},\Omega)}\mapsto{(\bar g_{ab},\bar
  \Omega)}:={(\Theta^2 g_{ab},\Theta\,\Omega)}$ with $\Theta>0$. All
the unphysical spacetimes $(M,\Theta^2 g_{ab}, \Theta\,\Omega)$ belong
to the same physical spacetime $(\tilde M,\tg_{ab})$.
\\
Under the rescaling $\bar g_{ab} = \Theta^2\,g_{ab}$, $R$ and $\bar R$
are connected by
\begin{equation}
\label{conformalGauge}
  6\, \nabla^a \nabla_a \Theta = \Theta R - \Theta^3 \bar R,
\end{equation}
which is equation (\ref{Rtrans}) where the covariant derivatives
$\nabla_a$ now corresponds to the unscaled metric. Solving
(\ref{conformalGauge}) for a spacetime ($M$, $g_{ab}$) and data
for $\Theta{}$ and $\dot \Theta{}$ on
a spacelike surface $S$ we get at least locally a unphysical
space time with arbitrary Ricci scalar $\bar{R}$.
\\
There is still conformal gauge freedom left as every rescaling with
$\Theta>0$ and
\begin{equation}
\label{ConformalGaugeFix}
  \nabla^a \nabla_a \Theta = \frac{1}{6} \Theta\,R\,\left(1-\Theta^2\right)
\end{equation}
leaves the Ricci scalar unchanged.
\\[2\parskip]
Equation (\ref{sfGl}) serves as regular equation for
$\Omega{}$. Substituting ${\omega = \frac{1}{4} \, \nabla^c \nabla_c
  \Omega}$ yields
\begin{equation}
\label{OmGl}
  \nabla_a \nabla_b \Omega = - \, \frac{1}{2} \, \Omega \, \hR_{ab}
                      + \omega \, g_{ab}
                      + \frac{1}{2} \, \Omega^3 \hT_{ab},
\end{equation}
which is a second order equation for $\Omega{}$.
\\
The next step is to find equations for the metric and the quantities
derived therefrom. Expressing the once contracted, second Bianchi
identity (${\nabla_{[a} R_{bc]d}{}^a =0}$)  in terms of $\hR_{ab}$ and
$d_{abc}{}^d$
results in
\begin{equation}
\label{hRGl}
  \nabla_{[a} \hR_{b]c} = - \, \frac{1}{12} \, (\nabla_{[a} R) g_{b]c}
                          - (\nabla_d \Omega) \, d_{abc}{}^d
                          - \Omega \, \nabla_d d_{abc}{}^d.
\end{equation}
The once contracted second Bianchi identity in the physical spacetime,
$$
  \tn_d \tC_{abc}{}^d =
  - \tn_{[a} (\tR_{b]c} - \frac{1}{6} \tg_{b]c} \tR),
$$
together with
$$
  \Omega^{-1} \tn_d \tC_{abc}{}^d = \nabla_d(\Omega^{-1}
                                    C_{abc}{}^d),
$$
and the Einstein equation in physical spacetime provide us with
an equation for $d_{abc}{}^d$:
\begin{eqnarray}
\label{dGl}
  \nabla_d d_{abc}{}^d
                       & = & - \, \Omega \, \nabla_{[a} \hT_{b]c}
                             - 3 \, (\nabla_{[a} \Omega) \, \hT_{b]c}
                             + g_{c[a} \hT_{b]d} \, (\nabla^d \Omega)
                                                        \nonumber  \\
                       &   & + \frac{1}{3} \, (\nabla_{[a} \Omega) \,
                       T \, g_{b]c}
                             + \frac{1}{12} \Omega \, (\nabla_{[a}T)
                               \, g_{b]c}  \; =: \;  t_{abc}.
\end{eqnarray}
We can now derive the missing equation for $\omega{}$ from the
integrability condition for (\ref{OmGl}) and by substituting (\ref{hRGl}):
\begin{eqnarray}
\label{omGl}
  \nabla_a \omega & = & - \frac{1}{2} \hR_{ab} \, \nabla^b \Omega
                    - \frac{1}{12} \, R \, \nabla_a \Omega
                    - \frac{1}{24} \Omega \, \nabla_a R
                    + \frac{1}{2} \Omega^2 \, \hT_{ab} \, \nabla^b \Omega
                                                       \nonumber   \\
                  &  &  - \frac{1}{6} \Omega^2 \, (\nabla_a \Omega) \, T
                        - \frac{1}{24} \Omega^3 \, \nabla_a T.
\end{eqnarray}
In the following in addition to the abstract indices (small Latin
letters) frame (underlined indices) and coordinate indices (Greek
letters) are used. The used conventions are explained in the appendix
in more detail.
\\
Using $\Omega_a := \nabla_a \Omega{}$, the frame  $e_{\f{i}}{}^a$, and
the Ricci rotation coefficients $\gamma^a{}_{\f{i}\f{j}}$ as further
variables, we get the following first order system of tensor equations
for  $\Omega{}$, $\Omega_a$, $\omega$, $e_{\f{i}}{}^a$,
$\gamma^a{}_{\f{i}\f{j}}$, $\hR_{ab}$, and $d_{abc}{}^d$: \footnote{
The symbol $E$ stands for equation, the first index reminds to the
quantity for which a temporary equation will be formed by setting the
tensor $E$ equal to $0$. The tensors providing the eventually used
forms of
the
equations are named with $\N{}$ standing for null quantities.}
\begin{subequations}
\label{quaSys}
\begin{eqnarray}
  \label{quaSysOm}
  \label{NO}
  \N{\Omega}_a = E\I{\Omega}_a & = & \nabla_a \Omega - \Omega_a = 0   \\
  \label{quaSysDOm}
  \label{NDO}
  \N{D\Omega}_{ab} = E\I{D\Omega}_{ab} & = &
    \nabla_a \Omega_b + \frac{1}{2} \Omega \hR_{ab}
    - \omega g_{ab} - \frac{1}{2} \Omega^3 \hT_{ab} = 0           \\
  \label{No}
  \N{\omega}_{ab} = E\I{\omega}_a & = &
    \nabla_a \omega + \frac{1}{2} \hR_{ab} \Omega^b
    + \frac{1}{12} R \Omega_a + \frac{1}{24} \Omega \nabla_a R
    - \frac{1}{2} \Omega^2 \hT_{ab} \Omega^b           \nonumber  \\
    && \qquad + \frac{1}{6} \Omega^2 \Omega_a T
    + \frac{1}{24} \Omega^3 \nabla_a T     =  0                   \\
  \label{quaSysDe}
  \label{Ne}
  \N{e}^a{}_{bc} = E\I{e}^a{}_{bc} & = & T^a{}_{bc}   = 0 \\
  \label{quaSysDgamma}
  \label{Ng}
  \N{\gamma}_{abc}{}^d = E\I{\gamma}_{abc}{}^d & = &
    R\I{diff}_{abc}{}^d - R\I{alg}_{abc}{}^d  = 0 \\
  \label{quaSysR}
  E\I{R}_{abc} & = & \nabla_{[a} \hR_{b]c}
    + \frac{1}{12} (\nabla_{[a} R) g_{b]c} + \Omega_d d_{abc}{}^d
    + \Omega \, t_{abc} = 0  \\                          
  \label{quaSysd}
  E\I{d}_{abc} & = &  \nabla_d d_{abc}{}^d  - t_{abc} = 0
\end{eqnarray}
\end{subequations}
where (\ref{quaSysDe}) means vanishing torsion $T^a{}_{bc}$, expressed
in frame index form,
\begin{equation*}
    T^{\f{i}}{}_{\f{j}\f{k}} =
    \left( e_{\f{j}}(e_{\f{k}}{}^\mu) -
           e_{\f{k}}(e_{\f{j}}{}^\mu) \right)
      e^{\f{i}}{}_\mu
  + \gamma^{\f{i}}{}_{\f{j}\f{k}} - \gamma^{\f{i}}{}_{\f{k}\f{j}},
\end{equation*}
and (\ref{quaSysDgamma}) means that the curvature tensor in terms of
the Ricci rotation coefficients, in frame index form
\begin{eqnarray*}
    R\I{diff}_{\f{i}\f{j}\f{k}}{}^{\f{l}} & = &
  e_{\f{j}}(\gamma^{\f{l}}{}_{\f{i}\f{k}})
    - e_{\f{i}}(\gamma^{\f{l}}{}_{\f{j}\f{k}})
  - \gamma^{\f{l}}{}_{\f{i}\f{m}} \gamma^{\f{m}}{}_{\f{j}\f{k}}
  + \gamma^{\f{l}}{}_{\f{j}\f{m}} \gamma^{\f{m}}{}_{\f{i}\f{k}}
  \nonumber  \\
  && \qquad
  + \gamma^{\f{m}}{}_{\f{i}\f{j}} \gamma^{\f{l}}{}_{\f{m}\f{k}}
  + \gamma^{\f{m}}{}_{\f{j}\f{i}} \gamma^{\f{l}}{}_{\f{m}\f{k}}
  - \gamma^{\f{l}}{}_{\f{m}\f{k}} T^{\f{m}}{}_{\f{j}\f{i}},
\end{eqnarray*}
should equal the combination
\begin{equation*}
  \Omega \, d_{abcd}
  + g_{c[a} \hR_{b]d} - g_{d[a} \hR_{b]c}
  + \frac{1}{6} g_{c[a} g_{b]d} R =: R\I{alg}_{abc}{}^d,
\end{equation*}
which is the irreducible decomposition of a tensor with the symmetry
of the Riemann tensor (\ref{Ralg}). Hence (\ref{quaSysDe}) and
(\ref{quaSysDgamma})
ensure that $R\I{alg}_{abc}{}^d$ is the curvature tensor corresponding
to the connection given by the Ricci rotation coefficients which again
is the torsion free connection coming from the metric (frame).
\subsubsection{Complications by the matter terms}
The final goal is to use the terms $\nabla_a \Omega{}$,  $\nabla_a
\Omega_b$, $\nabla_a \omega{}$, $\left( e_{\f{j}}(e_{\f{k}}{}^\mu) -
e_{\f{k}}(e_{\f{j}}{}^\mu) \right)  e^{\f{i}}{}_\mu{}$,\hfill
${e_{\f{j}}(\gamma^{\f{l}}{}_{\f{i}\f{k}})  -
e_{\f{i}}(\gamma^{\f{l}}{}_{\f{j}\f{k}})}$, $\nabla_{[a} \hR_{b]c}$, and
$\nabla_d d_{abc}{}^d$ in (\ref{quaSys}) as principal
part for the geometry variables of the system. I will call these terms
left side of the equations, the remaining terms right side. The left side
does not contain the complete principle part of the system yet as the
energy momentum
tensor $T_{ab}$ and its derivatives $\nabla_{[a} T_{b]c}$ may contain
derivatives of the matter and geometry variables.
\\
In the case of the conformally invariant scalar field the field equation
(\ref{model}) remains invariant under the rescaling
\begin{equation*}
  \phi = \Omega^{-1} \, \tp,
\end{equation*}
i.e.
\begin{equation*}
  {\vphantom{\phi}\Box} \phi - \frac{R}{6} \, \phi = 0.
\end{equation*}
The physical energy-momentum tensor $\tT_{ab}$ fulfills the assumed properties,
\begin{eqnarray*}
  \tT_{ab} & = &  \Om^2
               \left[
                  (\nabla_a \phi) (\nabla_b \phi)
                  - \frac{1}{2} \phi \nabla_a \nabla_b \phi
                  + \frac{1}{4} \phi^2 R_{ab}
                  - \frac{1}{4} g_{ab}
                  \left( (\nabla^c \phi) (\nabla_c \phi)
                         + \frac{1}{6} \phi^2 R \right)
               \right]  \nonumber \\
                   & =: &  \Om^2 \, T_{ab}.
\end{eqnarray*}
\\
The mentioned complications in (\ref{quaSys}) by the right sides are
now obvious. Firstly $\nabla_{[a} T_{b]c}$ contains $\nabla_{[a}
\nabla_{b]} \nabla_c \phi $ terms which are eliminated with the
identity $\nabla_{[a} \nabla_{b]} \nabla_c \phi{} = \frac{1}{2}
R_{abc}{}^d \nabla_d \phi $. To get rid of the second and first
order derivatives of $\phi $ we use the first order system
\begin{subequations}
\label{NM}
\begin{eqnarray}
\label{Np}
  \N{\phi}_a & = & \nabla_a \phi - \phi_a = 0
\\
\label{NDp}
  \N{D\phi}_{ab} & = &
  \nabla_a \phi_b - \hat{\phi}_{ab} - \frac{1}{4} \phi_c{}^c \, g_{ab} = 0
\\
\label{NBp}
  \N{\Box\phi} & = & \phi_a{}^a - \frac{R}{6} \, \phi = 0
\\
\label{NDDp}
  \N{DD\phi}_{abc} & = &
  \nabla _{[a} \hat{\phi}_{b]c}
  + \frac{1}{6} \, ( \phi\,\nabla_{[a} R + R \, \phi_{[a} ) g_{b]c}
  - \frac{1}{2} \, R\I{alg}_{abc}{}^d \phi_d = 0
\\
\label{NDBp}
  \N{D\Box\phi}_a & = &
  \nabla _{a} \phi_b{}^b
  - \frac{1}{6} \, ( \phi\, \nabla_a R + R \, \phi_{a} ) = 0.
\end{eqnarray}
\end{subequations}
for the variables $\phi$, $\phi_a$, the trace free symmetric tensor
$\hat{\phi}_{ab}$ and the trace $\phi_a{}^a$. The system is
derived from $\nabla_a
\left( {\vphantom{\phi}\Box} \phi - \frac{R}{6} \, \phi \right) =0$.
System (\ref{NM}) also serves as matter part of the system for the
unphysical spacetime.
$t_{abc}$ is now written in a form which does not contain any
derivatives of matter variables explicitly.
\\
$\nabla_{[a} T_{b]c}$ and thus $t_{abc}$ still contain
derivatives $\nabla_{[a} \hR_{b]c}$ of the trace free Ricci tensor.
By combining (\ref{quaSysR}) and (\ref{quaSysd}) the
derivatives of $\hR_{ab}$ and $d_{abc}{}^d$ can be decoupled.
(\ref{quaSysR}) and (\ref{quaSysd}) become
\begin{equation}
\label{quaSysRvar}
  E'\I{R}_{abc} = \nabla_{[a} \hR_{b]c}
    + \frac{1}{12} (\nabla_{[a} R) g_{b]c} - \Omega_d d_{abc}{}^d
    + \Omega m_{abc} = 0
\end{equation}
and
\begin{equation}
\label{quaSysdvar}
  E'\I{d}_{abc} = \nabla_d d_{abc}{}^d  - m_{abc} = 0,
\end{equation}
with
\begin{eqnarray*}
  \lefteqn{ m_{abc} = \frac{1}{1-\frac{1}{4} \, \Omega^2\phi^2} \qquad * }  \\
  \lefteqn{ \Big( \, \Omega } && \qquad
  \big[ \frac{3}{2}  \, \phi_{[a} \phi_{b]c}
    - \frac{1}{2}  \, g_{c[a} \phi_{b]d} \phi^d
    + \frac{1}{4}  \, \phi  \, \Omega d_{abc}{}^d \phi_d
    + \frac{1}{4}  \, \phi  \, g_{c[a} \hR_{b]}{}^d \phi_d
    - \frac{3}{4}  \, \phi  \, \phi_{[a} \hR_{b]c}        \\ && \qquad \quad
    - \frac{1}{12}  \, \phi  \, \phi_{[a} g_{b]c} R
    + \frac{1}{4}  \, \Omega  \, \phi^2 d_{abc}{}^d \Omega_d  \big]
\\ && \quad
  - 3  \, \Omega_{[a} \big[ \phi_{b]} \phi_c
    - \frac{1}{2}  \, \phi  \, \phi_{b]c}
    + \frac{1}{4}  \, \phi^2 \hR_{b]c}
    + \frac{1}{36}  \, \phi^2 g_{b]c} R
    - \frac{1}{3}  \, g_{b]c}  \, \phi^d \phi_d \big]
\\ && \quad
  + \Omega^d g_{c[a} \big[ \phi_{b]} \phi_d
    - \frac{1}{2}  \, \phi  \, \phi_{bd}
    + \frac{1}{4}  \, \phi^2 \hR_{b]d}    \big]     \quad     \Big).
\end{eqnarray*}
Note that $m_{abc}$ may become singular for
$1-\frac{1}{4}\,\Omega^2\phi^2= 1-\frac{1}{4}\,\tp^2=0$. In the
Einstein equations for the physical spacetime (\ref{EinstPhys})
$\tR_{ab}$ carries a factor $1-\frac{1}{4} \tp^2$ too.
We will need later that
\begin{equation*}
  \N{m}_{abc} := t_{abc} - m_{abc} =
  - \frac{1}{4}  \, \Omega  \, \phi^2
    \left( \N{R}_{abc} + \frac{2}{3} \, \Omega \, m_{[a|d|}{}^d \,
    g_{b]c}
\right),
\end{equation*}
where $\N{R}_{abc}$ is the null quantity representing the final form
of the equation for $\hR_{ab}$ (\ref{NR}).
The final form of  the equation for $d_{abc}{}^d$ is
obtained from (\ref{quaSysdvar}) by replacing $E'\I{d}_{abc}=0$ with
\begin{equation}
\label{Nd}
  \N{d}_{abc} := E'\I{d}_{abc} + \frac{2}{3} m_{[a|d|}{}^d g_{b]c} = 0.
\end{equation}
This gives $\N{d}_{abc}$ the same index symmetry properties as the Weyl tensor.
That replacement does not change the equation
since $m_{ab}{}^b=0$ as will be seen later.
\\
Analogously we replace (\ref{quaSysRvar}) with
\begin{equation}
\label{NR}
  \N{R}_{abc} :=
    E'\I{R}_{abc} - \frac{2}{3} \, \Omega \, m_{[a|d|}{}^d g_{b]c} = 0,
\end{equation}
the contraction $\N{R}_{ab}{}^b=0$ is then the contracted second
Bianchi identity.
\section{Evolution equations and constraints}
In the following I will assume a system $\N{}=0$ of the form (\ref{quaSysOm})
-- (\ref{quaSysDgamma}),  (\ref{Nd}), (\ref{NR}),  the geometry part,
and a matter part, in the case of the scalar field model system
(\ref{NM}). $m_{abc}$ and $t_{abc}$ are assumed to differ only by terms
expressible as null quantities. The energy-momentum tensor
$T_{ab}$, its derivatives $\nabla_a T_{bc}$, $m_{abc}$, and $t_{abc}$
are assumed to be expressed in variables and thus do not contain any
explicit derivative of variables. By these assumptions the
principal part of the system has block form, the geometry block and
the matter block. The two blocks are coupled through the right sides.
\\
In this chapter the system $\N{}=0$ will be reduced to a system of
symmetric hyperbolic time evolution equations, the subsidiary system.
Sufficient conditions for the equivalence of the subsidiary system and
$\N{}=0$ are given as conditions on $m_{abc}$. If the system can be put
into the described block form there do not arise any more conditions
from the geometry part of the system for any matter.
\\
The explicit carry out is technical and lengthy, the idea can be
summarized as follows: All the equations of the system are
regarded as null quantities. By requiring the vanishing of some of these null
quantities and by choosing an appropriate gauge condition for the
coordinates and the frame we get a symmetric hyperbolic
subsidiary system of evolution equations.
Which null quantities to choose can best be seen by a decomposition into
the irreducible parts in the spinor calculus as performed
in~\cite{Fr91ot}.
\\
The solution of this symmetric hyperbolic subsystem exists and is
unique. To complete the proof we must show
that the solution obtained in this way is consistent with the rest of
the equations, i.e.\ that all null quantities remain zero if they are
initially zero (``propagation of the constraints''). For that purpose a
symmetric hyperbolic system of time evolution equations for
the remaining null quantities is derived. Sufficient conditions for the
propagation of the constraints are firstly the homogeneity of the evolution
equations for the remaining null quantities in the null quantities
since then the unique solution of these evolution equations is the vanishing
of all null quantities for all times if they vanish on the initial
surface and secondly that the domain of dependence of $S$ with respect
to the equations for the propagation of the constraints is a
superset of the domain of dependence of $S$ with respect to the
subsidiary system.
\subsection{A symmetric hyperbolic subsystem of evolution
  equations}
\label{GaussGauge}
Introducing a timelike vector field $t^a$ not necessarily hypersurface
orthogonal and its orthogonal projection tensor
$h_{ab}:=g_{ab}-t_at_b/(t_ct^c)$ allows to split the
system of equations into two categories, the equations containing time
derivatives and those containing no time derivatives (the
constraints). The equations with time derivatives provide a
under/overdetermined system of evolution equations.
\\
The system is
overdetermined since for some quantities there are too many time
evolution equations, e.g.\ there are 12 time evolution equations from
$\N{R}_{abc}=0$
for 9 independent tensor components. 3 equations are a linear
combination of the other 9 equations and the constraints. An
irreducible decomposition of the tensors $\N{}$ is a systematic way to
analyze these dependencies. Since all the types of tensor index
symmetries appearing in the system have been thoroughly investigated
in~\cite{Fr91ot} I will only state which combinations are needed.
\\
The system is underdetermined since there are 10 time evolution
equations for the frame and the Ricci rotation coefficients missing.
By adding
\begin{equation}
\label{KoEichFrame}
   e^{\f{i}}{}_b \, g^{\f{j}\f{k}} \, e_{\f{j}}{}^a \left( \nabla_a
   e_{\f{k}}{}^b \right) = - F_{\f{i}} =
   \gamma_{\f{i}\f{k}}{}^{\f{k}}
\end{equation}
and
\begin{equation}
\label{FrEichFrame}
    \partial_{\f{k}} \gamma^{\f{i}\f{k}\f{j}}
    + \gamma^{\f{i}\f{k}\f{j}} F_{\f{k}}
    + \gamma^{\f{l}\f{k}\f{i}} \, \gamma_{\f{l}\f{k}}{}^{\f{j}}
    - \gamma^{\f{l}\f{k}\f{j}} \, \gamma_{\f{l}\f{k}}{}^{\f{i}}
    = F^{\f{i}\f{j}},
\end{equation}
the system becomes complete. The freedom of giving ten functions
corresponds to the freedom of giving the lapse and the shift to
determine the coordinates and the six parameters of the Lorentz group
to determine the frame on every point. The gauge freedom is discussed
in full detail in~\cite{Fr85ot,Fr91ot}.
\\
A choice which makes the system especially simple for analytic
considerations is a Gaussian coordinate and frame system defined as
follows. Give on the spacelike initial value surface $S$ coordinates
$x^\mu, \mu = 1..3$, and 3 orthonormal vector fields $e_{\f{i}}{}^a,
\f{i}=\f{1}..\f{3}$ in this hypersurface. The affine parameter of the
geodesics of the hypersurface orthonormal, timelike vector field
$e_{\f{0}}{}^a$ defines the
time coordinate $x^0=t$. The spacelike coordinates are transported by
these geodesics into a neighbourhood of the initial surface. By
geodesic transport of $e_{\f{i}}{}^a, \f{i}=\f{1}\ldots\f{3}$, and
$e_{\f{0}}{}^a$  a frame is obtained in this neighbourhood. By
construction we have
$$
  e_{\f{0}}{}^0=1, \quad e_{\f{0}}{}^\mu = 0 \mbox{ for }\mu = 1..3
$$
and
$$
  \gamma^{\f{i}}{}_{\f{0}\f{k}} = 0.
$$
It is well known that Gaussian coordinates develop caustics if the
energy momentum tensor fulfills certain energy conditions, see
e.g.~\cite[lemma 9.2.1]{Wa84GR}. In the
unphysical spacetime the $\Omega{}$ terms provide a kind of unphysical
energy-momentum tensor.
Whether this energy-momentum tensor fulfills the energy-momentum
conditions is a difficult question and not known to the author.
Nevertheless the coordinates develop caustics as has been shown by
numerical calculations \cite{Hu93nu}.
\\
The following combinations give a symmetric hyperbolic
system for the remaining variables as can be deduced from the
considerations in~\cite{Fr83cp}:
\begin{subequations}
\label{EvoSyst}
\begin{eqnarray}
&& \N{\Omega}_{\f{0}} = 0 , \\
&& \N{D\Omega}_{\f{0}b} = 0 , \\
&& \N{\omega}_{\f{0}} = 0 , \\
&& \N{e}^a{}_{b\f{0}} = 0 , \\
&& \N{\gamma}_{\f{0}\f{1}c}{}^d = 0 , \\
&& g^{ab} \N{R}_{\f{i}ab} = 0 , \qquad \f{i} = \f{1},\f{2},\f{3}, \\
&& \N{R}_{\f{0}\f{i}\f{i}} = 0 , \qquad \f{i} = \f{1},\f{2},\f{3}, \\
&& \N{R}_{\f{0}\f{i}\f{j}} + \N{R}_{\f{0}\f{j}\f{i}} = 0 ,
   \qquad (\f{i},\f{j}) = (\f{1},\f{2}),(\f{1},\f{3}),(\f{2},\f{3}), \\
&& \N{d}_{\f{2}\f{1}\f{2}} - \N{d}_{\f{3}\f{1}\f{3}} +
   \N{d}_{\f{2}\f{0}\f{2}} - \N{d}_{\f{3}\f{0}\f{3}} = 0 , \\
&& - \N{d}_{\f{1}\f{0}\f{2}} + \N{d}_{\f{1}\f{2}\f{1}} = 0 , \\
&& \N{d}_{\f{1}\f{0}\f{1}} = 0 , \\
&& \N{d}_{\f{1}\f{0}\f{2}} + \N{d}_{\f{1}\f{2}\f{1}} = 0 , \\
&& - \N{d}_{\f{2}\f{1}\f{2}} +
   \N{d}_{\f{3}\f{1}\f{3}} + \N{d}_{\f{2}\f{0}\f{2}} -
   \N{d}_{\f{3}\f{0}\f{3}} = 0 , \\
&& \N{d}_{\f{2}\f{1}\f{3}} + \N{d}_{\f{3}\f{1}\f{2}} + \N{d}_{\f{2}\f{0}\f{3}}+
   \N{d}_{\f{3}\f{0}\f{2}} = 0 , \\
&& - \N{d}_{\f{1}\f{0}\f{3}} + \N{d}_{\f{1}\f{3}\f{1}} = 0 , \\
&& - \N{d}_{\f{1}\f{2}\f{3}} = 0 , \\
&& - \N{d}_{\f{1}\f{0}\f{3}} - \N{d}_{\f{1}\f{3}\f{1}} = 0 , \\
&& \N{d}_{\f{2}\f{1}\f{3}} + \N{d}_{\f{3}\f{1}\f{2}} -
   \N{d}_{\f{2}\f{0}\f{3}} - \N{d}_{\f{3}\f{0}\f{2}} = 0 , \\
&& \N{\phi}_{\f{0}} = 0 , \\
&& \N{D\phi}_{\f{0}b} = 0 , \\
&& g^{ab} \N{DD\phi}_{\f{i}ab} = 0 , \qquad \f{i} = \f{1},\f{2},\f{3},
\\
&& \N{DD\phi}_{\f{0}\f{i}\f{i}} = 0 , \qquad \f{i} = \f{1},\f{2},\f{3},
\\
&& \N{DD\phi}_{\f{0}\f{i}\f{j}} + \N{DD\phi}_{\f{0}\f{j}\f{i}} = 0
, \qquad (\f{i},\f{j}) = (\f{1},\f{2}),(\f{1},\f{3}),(\f{2},\f{3}), \\
&& \N{D\Box\phi}_{\f{0}} = 0.  \yesnumber
\end{eqnarray}
\end{subequations}
To see that the system is really symmetric hyperbolic one has to write
down the system explicitly. By an appropriate, in the explicit
form of the system obvious definition of new variables, the system has
the structure
\begin{equation}
\label{symhypEvoSys}
  \underline{\underline{A_t}}\,\partial_t \underline{f} +
  \sum_{i=1}^3 \underline{\underline{A_{x^i}}}\,\partial_{x^i}
  \underline{f} +
  \underline{b}(\underline{f},x^\mu) = 0,
\end{equation}
with a diagonal matrix
$\underline{\underline{A_t}}$, which is positive definite for
$1-\frac{1}{4}\Omega^2\phi^2>0$, and symmetric matrices
$\underline{\underline{A_{x^i}}}$. $\underline{f}$ is the vector build
from the variables.
All the remaining equations are linear combinations of~(\ref{EvoSyst})
and constraints.  Since the explicit form of the constraints is not
needed I do not list them.
\\
As the entries in $\underline{\underline{A_t}}$ coming from
$\N{R}=0$ and $\N{d}=0$ vanish for $1-\frac{1}{4}\Omega^2\phi^2=0$ the
following results apply only if $\Omega^2\phi^2<4$ everywhere on the
initial value surface $S$ and thus in a neighbourhood of $S$. The
physical Einstein equations have a corresponding singularity for
$1-\frac{1}{4}\tp^2=0$ (see equation \ref{EinstPhys}).
\subsection{A sufficient condition for the propagation of the
  constraints}
According to the analysis in~\cite{Fr91ot}, involving the left hand
side of the following identities, a symmetric
hyperbolic system of evolution equations for the remaining null quantities
can be extracted from:
\begin{subequations}
\label{PropConstrSys}
\begin{equation}
  \nabla_{[a}\N{\Omega}_{b]} =
  - \frac{1}{2} T^c{}_{ab} \nabla_c \Omega - \N{D\Omega}_{[ab]}
\ee

\begin{eqnarray}
  \lefteqn{ \nabla_{[a}\N{D\Omega}_{b]c} = }   \nonumber   \\ && \qquad
  \frac{1}{2} \N{\gamma}_{abc}{}^d \Omega_d
  - \frac{1}{2} T^d{}_{ab} \nabla_d \Omega_c
  + \frac{1}{2} \Omega \N{R}_{abc}
  + \frac{1}{2} \hR_{c[b} \N{\Omega}_{a]}
  - \frac{3}{2} \Omega^2 \N{\Omega}_{[a} \hT_{b]c} \nonumber \\ && \qquad
  - \frac{1}{2} \N{\omega}_{[a} g_{b]c}
  + \frac{1}{3} \Omega^2 m_{[a|d|}{}^d g_{b]c}
  + \frac{1}{3} \Omega^2 \N{m}_{[a|d|}{}^d g_{b]c}
\end{eqnarray}

\begin{eqnarray}
  \lefteqn{ \nabla_{[a}\N{\omega}_{b]} = }  \nonumber          \\ && \qquad
  - \frac{1}{2} T^c{}_{ab} \nabla_c \omega
  - \frac{1}{24} \Omega^3 T^c{}_{ab} \nabla_c T
  + \frac{1}{24} \N{\Omega}_{[a} \nabla_{b]} R
  + \frac{1}{8} \Omega^2 \N{\Omega}_{[a} \nabla_{b]} T  \nonumber \\ && \qquad
  + \frac{1}{2} ( \hR _{c[b} - \Omega^2 \hT_{c[b} ) \N{D\Omega}_{a]}{}^c
  + ( \frac{1}{24} R + \frac{1}{6} \Omega^2 T ) \N{D\Omega}_{[ab]}
  + \frac{1}{2} \Omega^c \N{R}_{abc}             \nonumber        \\ && \qquad
  + \frac{1}{3} \Omega \, \Omega_{[b} m_{a]c}{}^c
  + \frac{1}{2} \Omega \, \Omega^c \N{m}_{abc}
\end{eqnarray}

\begin{eqnarray}
\label{PropNR}
  \lefteqn{ \nabla_{[a} \N{R}_{bc]d} = }        \nonumber  \\ && \qquad
  \frac{1}{2} \N{\gamma}_{[abc]}{}^f \hR_{fd}
  + \frac{1}{2} \N{\gamma}_{[ab|d]}{}^f \hR_{c]f}
  - \frac{1}{2} T^f{}_{[ab} \nabla_{|f|}\hR_{c]d}
  - \frac{1}{24} T^f{}_{[ab} (\nabla_{|f|} R ) g_{c]d} \nonumber \\ && \qquad
  + \N{D\Omega}_{[a|f|} d_{bc]d}{}^f
  + \Omega^f ( \N{d}_{[ca|d|} g_{b]f} - \N{d}_{[ca|f|} g_{b]d} )
  + \N{\Omega}_{[a} m_{bc]d}           \nonumber                 \\ && \qquad
  - \frac{2}{3} \N{\Omega}_{[a} m_{b|f|}{}^f g_{c]d}
  - \frac{1}{2} \Omega^2 \N{\gamma}_{[abc]}{}^f \hT_{fd}
  - \frac{1}{2} \Omega^2 \N{\gamma}_{[ab|d|}{}^f \hT_{c]f}
  - \frac{1}{2} \Omega^2 T^f{}_{[ab} \nabla_{|f|} \hT_{c]d} \nonumber
  \\ &&
\qquad
  - 3 \Omega \N{D\Omega}_{[ab} \hT_{c]d}
  + \Omega \N{D\Omega}_{[a}{}^f \hT_{c|f|} g_{b]d}
  + \frac{1}{3} \Omega \N{D\Omega}_{[ab} g_{c]d}
  + \frac{1}{12} \Omega \N{\Omega}_{[a} ( \nabla_b T ) g_{c]d}
    \nonumber  \\ && \qquad
  + \frac{1}{12} \Omega^2 T^f{}_{[ab} ( \nabla_{|f|} T ) g_{c]d}
  - 2 \Omega_{[a} \N{m}_{bc]d}
  + \Omega_f \N{m}_{[ca}{}^f g_{b]d}      \nonumber              \\ && \qquad
  - \Omega \nabla_{[a} \N{m}_{bc]d}
  + 2 \Omega_{[a} m_{c|f|}{}^f g_{b]d}
  - \frac{2}{3} \Omega ( \nabla_{[a} m_{b|f|}{}^f ) g_{c]d}
\end{eqnarray}

\begin{eqnarray}
  \lefteqn{ \nabla^{c} \N{d}_{abc} = }  \nonumber     \\ && \qquad
  \frac{1}{2} \N{\gamma}^c{}_{da}{}^f d_{fbc}{}^d
  + \frac{1}{2} \N{\gamma}^c{}_{db}{}^f d_{afc}{}^d
  + \frac{1}{2} \N{\gamma}^c{}_{dc}{}^f d_{abf}{}^d
  + \frac{1}{2} \N{\gamma}^{cd}{}_{df} d_{abc}{}^f  \nonumber   \\ && \qquad
  - \frac{1}{2} T^{fcd} \nabla_f d_{abcd}
  - \nabla^c m_{abc}
  + \frac{2}{3} \nabla_{[b} m_{a]c}{}^c
\end{eqnarray}

\begin{equation}
  \nabla_{[a} \N{e}^d{}_{bc]} =
  \N{\gamma}_{[abc]}{}^d + \N{e}^f{}_{[ab} \N{e}^d{}_{c]f}
\ee

\begin{eqnarray}
\label{dNgamma}
  \lefteqn{ \nabla_{[f} \N{\gamma}_{ab]cd} = } \nonumber \\ && \qquad
  - T^g{}_{[fa} R\I{diff}_{b]gcd}
  - \N{\Omega}_{[f} d_{ab]cd}
  - \Omega ( \N{d}_{[bf|c|} g_{a]d} - \N{d}_{[bf|d|} g_{a]c} )
  \nonumber \\ && \qquad
  - \N{R}_{[fb|d|} g_{a]c}
  + \N{R}_{[fb|c|} g_{a]d}
\end{eqnarray}

\begin{equation}
  \nabla_{[a}\N{\phi}_{b]} =
  - \frac{1}{2} T^c{}_{ab} \nabla_c \phi - \N{D\phi}_{[ab]}
\ee

\begin{equation}
  \nabla_{[a}\N{D\phi}_{b]c} =
  \frac{1}{2} \N{\gamma}_{abc}{}^d \phi_d
  - \frac{1}{2} T^d{}_{ab} \nabla_d \phi_c
  - \N{DD\phi}_{abc}
  - \frac{1}{4} \N{D\Box\phi}_{[a} g_{b]c}
\end{equation}

\begin{equation}
  \nabla_a \N{\Box\phi} =
  \N{D\Box\phi}_a
  - \frac{1}{6} R \N{\phi}_a
\end{equation}

\begin{eqnarray}
\label{PropNDDp}
  \lefteqn{ \nabla_{[a} \N{DD\phi}_{bc]d} = }   \nonumber  \\ && \qquad
  \frac{1}{2} \N{\gamma}_{abc}{}^f \hat{\phi}_{fd}
  + \frac{1}{2} \N{\gamma}_{[ab|d|}{}^f \hat{\phi}_{c]f}
  - \frac{1}{2} T^f{}_{[ab} \nabla_{|f|} \hat{\phi}_{c]d}
  + \frac{1}{6} \N{\phi}_{[a} ( \nabla_b R ) g_{c]d}   \nonumber \\ && \qquad
  - \frac{1}{12} \phi T^f{}_{[ab} g_{c]d} \nabla_f R
  + \frac{1}{6} R \N{D\phi}_{[ab} g_{c]d}
  + \frac{1}{2} T^g{}_{[ab} R\I{diff}_{c]gd}{}^f \phi_f
  - \frac{1}{2} R_{[bc|d|}{}^f \N{D\phi}_{a]f}        \nonumber \\ && \qquad
  + \frac{1}{2} ( \nabla_{[a} \N{\gamma}_{bc]d}{}^f ) \phi_f,
\end{eqnarray}
and
\begin{equation}
  \nabla_{[a} \N{D\Box\phi}_{b]} =
  - \frac{1}{2} T^d{}_{ab} \nabla_d \phi_c{}^c
  - \frac{1}{6} \N{\phi}_{[a} \nabla_{b]} R
  - \frac{1}{12} \phi T^c{}_{ab} \nabla_c R
  - \frac{1}{6} R \N{D\phi}_{[ab]}.
\end{equation}
\end{subequations}
The last term in (\ref{PropNDDp}) is homogeneous in null quantities as
can be seen from (\ref{dNgamma}),
The deviation of these equalities is even more lengthy than the
equalities itself, but the essential ideas behind it can already be seen
in the deviation of the first:
\begin{eqnarray*}
  \nabla_{[a}\N{\Omega}_{b]} & = &
    \nabla_{[a}  \nabla_{b]}\Omega - \nabla_{[a}  \Omega_{b]} \\
  & = &
    - \frac{1}{2} T^c{}_{ab} \nabla_c \Omega - \N{D\Omega}_{[ab]}
    + \frac{1}{2} \hR_{[ab]} \Omega
    - \frac{1}{2} \Omega^3 \hT_{[ab]}
    - \omega g_{[ab]},
\end{eqnarray*}
with the last three terms vanishing since the tensors are symmetric. Note
that vanishing of the torsion, $T^c{}_{ab}=0$,  and
$2\,\nabla_{[a}\nabla_{b]}\omega_c=R_{abc}{}^d \omega_d$ cannot be
used since they only
hold if both the time evolution and the constraint equations for the frame
$e_{\f{i}}{}^\mu $ and the Ricci rotation coefficients
$\gamma^a{}_{\f{i}\f{j}}$ hold everywhere.
\\
A sufficient set of conditions for homogeneity of the system derived
in the null quantities is
\begin{subequations}
\label{IntegrBed}
\begin{equation}
\label{IntegrBed1}
  m_{ab}{}^b = 0 \bmod \N{},
\ee
\begin{equation}
\label{IntegrBed2}
  \nabla^c m_{abc} + \frac{2}{3} \nabla_{[a} m_{b]c}{}^c = 0 \bmod \N{},
\ee
\begin{equation}
\label{IntegrBed3}
  \nabla_{[a} m_{b]c}{}^c = 0 \bmod \N{},
\ee
and
\begin{equation}
\label{IntegrBed4}
  \Omega \nabla_{[a} \N{m}_{bc]d} =
     f \, \nabla_{[a} \N{R}_{bc]d} \bmod \N{}, \quad f \ne -1.
\ee
\end{subequations}
A straightforward but long calculation shows that these conditions are
fulfilled by the conformally invariant scalar field with
$f=-\frac{1}{4}\Omega^2\phi^2$. Equation  (\ref{PropNR})
becomes singular for $1-\frac{1}{4}\Omega^2\phi^2=0$.
\\
The very technical integrability conditions (\ref{IntegrBed}) have a
very simply interpretation. Replacing $m_{abc}$ with $t_{abc}$ --- they
only differ by null quantities --- the conditions
(\ref{IntegrBed1}--\ref{IntegrBed3}) reduce to $\tn^b \tT_{ab} = 0$
and ${\tn_b \tn^c \tT_{ac} = 0}$. Condition (\ref{IntegrBed4}) is only
of technical nature, it gives the principal part a
simple block form.
\\
{}From the considerations in \cite{Fr91ot} also follows that the domain
of dependence of $S$ with respect to the evolution equation of the
constraints includes the domain of dependence of $S$ with respect to
the subsidiary system.
\section{The hyperboloidal initial value problem}
\label{HypInitValProblSec}
So far a system of equations $(\N{}=0)$ has been derived which contains
for at least one choice of gauge a symmetric hyperbolic subsystem of
evolution equations. The remaining equations in the system --- either
constraints or a combination of constraints and time evolution
equations --- will be satisfied for a solution of the evolution
equations, if the constraints are satisfied by the initial data. If
both, the time evolution and the constraints, are fulfilled, $(\tilde
M,\tg_{ab},\tp)$ is a weakly asymptotically flat solution of the Einstein
equation. This follows from the way the system $\N{}=0$ for the
unphysical spacetime has been derived.
\\
The essential points in the proofs of the theorems in
\cite[chapter~10]{Fr91ot} are the symmetric hyperbolicity of the
subsidiary system and the form (\ref{NDO}) of the equations for
$\Omega{}$. Therefore the same techniques can be used and the proofs
will not be repeated. The difference to
the model treated here lies in the derivation of the subsidiary system
and the proof of the propagation of the constraints,
which has been done in the previous chapters.
\subsection{The initial value problem}
We consider the following initial value problem:
\begin{Def}
\label{HypInitValProbl}
A ``{\bf hyperboloidal initial data set for the conformally invariant
scalar field}'' consists of a pair $(\bar S,f_0)$ such that:
\begin{enumerate}
\item $\bar S = S \cap \partial S$ is a smooth manifold with boundary
  $\partial S$ diffeomorphic to the closed unit ball in $\Bbb{R}^3$.
  As coordinates on $S$ the pull backs of the natural coordinates
  on $\Bbb{R}^3$ are used.
\item $f_0$ is the vector $\underline f$ of functions in system
  (\ref{EvoSyst}) written in the form (\ref{symhypEvoSys}) at initial
  time $t_0$.
\item The fields provided by $f_0$ have uniformly continuous derivatives
  with respect to the coordinates of $S$ to all orders\footnote{The
    assumption about the smoothness of the data can certainly be
    weakened from $C^\infty{}$ to $C^n$ for sufficiently large $n$ but
    then more technical effort would be needed in the proofs.}.
\item On $S$: $\Omega>0$. On $\partial S$: $\Omega=0$ and $\nabla_a
  \Omega{}$ is a future directed null vector.
\item The fields provided by $f_0$ satisfy the constraints following
  from $\N{}=0$ (\ref{quaSys} and \ref{NM}) and the gauge conditions.
\end{enumerate}
\end{Def}
A point which deserves special notice is the existence of a
hyperboloidal initial data set. The proof that
those data exist has to overcome two problems.
\\
Firstly,  the regularity of the solution on $\partial S$ which is
the consistency of the data with asymptotical flatness.
For scalar field data with compact support regularity
conditions are given in~\cite{AnCXXxx,AnCA92ot}, which are sufficient
for the existence of a solution of the constraints near $\partial S$.
\\
Secondly there is a problem with a possible singularity of
equations in $\N{}=0$ at {$1-\frac{1}{4}\,\Omega^2\phi^2=0$}.
\\
L.~Anderson and
P.~Chrusc\'{\i}el are preparing a paper analyzing both
problems \cite{AnCXXxx}.
\subsection{Theorems}
The ``theorems'' will be given in a form not containing every
technical detail, since these technical details would make them
lengthy and
can
be easily deduced from the theorems in~\cite{Fr91ot} by replacing the
Yang-Mills matter with the (conformally invariant) scalar field.
\\
Since the constraints of $\N{}=0$ propagate we have:
\begin{Theorem}
\label{physUnphysequi}
  Any (sufficiently smooth) solution of the subsidiary system satisfying the
  constraints on a spacelike hypersurface $\bar S$ and
  $1-\frac{1}{4}\,\Omega^2\phi^2>0$  defines in the
  domain of dependence with respect to $g_{ab}$ of $\bar S$ a
  solution to the unphysical system. Thus $(\tilde M,\tg_{ab},\tp)$ is
  a weakly asymptotically flat solution of the Einstein equation.
\end{Theorem}
Since the evolution equations are symmetric hyperbolic a unique
solution of the initial value problem exists for a finite time.
{}From the combination with theorem (\ref{physUnphysequi}) follows:
\begin{corollar}
\label{ExUni}
  For every regular solution of the constraints
  on $\bar S$ with ${1-\frac{1}{4}\,\Omega^2\phi^2\mid_{\bar S} >0}$ exists
  locally a unique, weakly asymptotically
  flat solution of the Einstein equation.
\end{corollar}
For the Minkowski space we can extent $\bar S$ and the solution of
the constraints beyond $\partial S$ to $S'$ and
get a solution in the unphysical spacetime which extents beyond
$i^+$. The continuous dependence of the solution of symmetric
hyperbolic systems on the data and the form of (\ref{NO}),
(\ref{NDO}) and (\ref{No}) (see the proof
of theorem (10.2) in~\cite{Fr91ot}) guarantees
that there is a solution covering the whole domain of dependence of
$\bar S$. Furthermore the proof there shows that
$\{p\,|\,\Omega(p)=0\}$ has an isolated critical point $i^+$, where
all future directed timelike geodesics of $(\tilde M,\tg_{ab})$ end, thus:
\begin{Theorem}
  For a sufficient small deviation of the data from Minkowskian data the
  solution of theorem~\ref{ExUni} possesses a regular future null
  infinity and a regular future timelike infinity.
\end{Theorem}
\section{The conformal equivalence of the scalar fields}
\label{SkalarAequiv}
This section shortly reviews the equivalence transformation between
spacetime models with scalar matter under the viewpoint of solving
hyperboloidal initial value problems. Other
aspects of this equivalence transformation, especially the generation
of exact solutions, have been studied
in~\cite{AcWA93ce,Be74es,Be75bh,Kl93sf,KlK93ie,Pa91mw,XaD92eg}.
\subsection{Local equivalence of solutions}
\label{GenEquivalence}
Spacetime models $(\tilde M,\tilde g_{ab},\tilde\phi)$
with scalar matter $\tilde\phi$ described by the action
\begin{equation}
\label{ScalarAction}
  \tilde S = \int_{\tilde M} \left[ A(\tilde\phi) R - B(\tilde\phi)
                    (\tilde\nabla_a\tilde\phi)
                    (\tilde\nabla^a\tilde\phi)
             \right] \> (-\tilde g)^{\frac{1}{2}} \> d^4\tilde x
\end{equation}
will be considered. Boundary terms in the action have been omitted,
$\tg{}$ is the determinant of $\tg_{\mu\nu}$.
\\
By varying the action $\tilde S$ with respect to $\tp{}$ and
$\tg_{ab}$ the following field equations result:
\begin{subequations}
\label{allgSys}
\begin{eqnarray}
\label{allgWell}
  B(\tp) \, \tilde{\vphantom{\phi}\Box} \,\tilde\phi
  + \frac{1}{2} \, \frac{dB}{d\tilde\phi} \,
    \left( \tn^a\tilde\phi \right) \, \left( \tn_a\tilde\phi \right)
  + \frac{1}{2} \, \frac{dA}{d\tilde\phi} \, \tilde R
  \quad & =  & \quad 0 \qquad \qquad
\\
  A(\tilde\phi) \, \left( \tilde R_{ab} - \frac{1}{2} \, \tilde R
  \, \tilde g_{ab} \right)
  + B(\tilde \phi) \, \left( \frac{1}{2} \,
    \left(\tn^c\tilde \phi\right) \,
               \left(\tn_c\tilde \phi\right) \tg_{ab}
    - \left(\tn_a\tilde \phi\right) \,
       \left(\tn_b\tilde \phi\right) \right)
\nonumber
\\
\label{allgGeo}
  - \left( \tn_a\tn_b A(\tilde\phi) \right)
   + \left( \tn^c\tn_c A(\tilde\phi) \right) \tilde g_{ab}
  \quad & = & \quad 0.
\end{eqnarray}
\end{subequations}
$A$ and $B$ are assumed to be $C^\infty$ functions.
For $ B \neq 0$ the principal part of (\ref{allgWell}) does not vanish
and thus (\ref{allgWell}) is a wave equation. For that reason I assume
$B(\tilde\phi) > \epsilon > 0$ for every $\tilde\phi$. (\ref{allgGeo})
is a second order equation for the metric if $A(\tp)\ne{}0$.
\\
In the spacetime region $\tilde H := \left\{x \in \tilde{M} | \, {\rm
    sign}(A(\tilde\Phi))>0 \quad \forall
  \tilde\Phi \in [\tp_0,\tp(x)]\right\}$ the trans\-for\-ma\-tion\footnote{The
    choice of the parameter $\tp_0$ reflects gauge freedom. Models
    where there is no $\tp_0$ with $A(\tp_0) > 0$ will not be
    considered.}
\begin{subequations}
\label{Transf}
\begin{eqnarray}
\label{allgphiTrans}
  \tilde{\bar\phi} & = & \int_{\tilde\phi_0}^{\tilde\phi} \frac{1}{A}
    \sqrt{ \frac{3}{2} \,
    \left(\frac{dA}{d\phi} \right)^2 + A\,B } \quad
  d\phi
\\
\label{allgTrans}
  \tilde {\bar g}_{ab} & = & A \, \tilde g_{ab}
\end{eqnarray}
\end{subequations}
gives a solution of the system (\ref{allgSys}) with a massless
Klein-Gordon field
$\widetilde{\bar\phi}$ as matter model corresponding to the choice
$(A,B)=(1,1)$ and the equations
\begin{subequations}
\label{KGGl}
\begin{eqnarray}
  \widetilde{\bar{\vphantom{\phi}\Box}}\widetilde{\bar\phi} & = & 0
\\
  \widetilde{\bar{R}}_{ab} - \frac{1}{2} \, \widetilde{\bar{R}} \,
  \widetilde{\bar{g}}_{ab}
  & = & \widetilde{\bar{T}}_{ab}[{\widetilde{\bar{\phi}}}]
\end{eqnarray}
with energy momentum tensor
\begin{equation}
  \widetilde{\bar{T}}_{ab}[{\widetilde{\bar{\phi}}}] =
  (\widetilde{\bar\nabla}_a\widetilde{\bar{\phi}}) \,
  (\widetilde{\bar\nabla}_b\widetilde{\bar{\phi}})
  - \frac{1}{2} \, (\widetilde{\bar\nabla}_c
                          \widetilde{\bar{\phi}}) \,
                         (\widetilde{\bar\nabla}^c
                          \widetilde{\bar{\phi}})
                          \, \widetilde{\bar g}_{ab}.
\end{equation}
\end{subequations}
\\
{}From the assumptions about $A$ and $B$ follows that the corresponding
Klein-Gordon field will be unbounded approaching the
part of the boundary of $\tilde H$ where $A(\tp)\rightarrow 0$.
The singularity in the Klein-Gordon field
shows up at least in a singularity of the equations for $(\tilde M,\tilde
g_{ab}, \tilde \phi)$.
\\
For two of the scalar fields in the above class the field equations
are very special, the already mention massless Klein-Gordon field
$\tilde{\bar\phi{}}$
(\ref{KGGl}) and the conformally invariant scalar field $\tp{}$,
$(A,B)=(1-\frac{1}{4}\tp^2,\frac{3}{2})$ ($\tp{}$ can be
rescaled by an arbitrary factor).
\\
The first, because the set of equations in the physical spacetime
becomes remarkable simple and has been analyzed intensely with
analytical (e.g.~\cite{Ch91tf}) and numerical
(e.g.~\cite{Ch92CB}) methods for spacetimes with spherical symmetry.
\\
The second, yielding the equations (\ref{model}),
because the matter equations are invariant under rescalings $g_{ab} =
\Omega^2 \, \tg_{ab}$ and $\phi = \Omega^{-1} \tp{}$.
\\
The transformation between the two special cases is
\begin{subequations}
\begin{eqnarray}
  \tilde{\B{\phi}}    & = &
    \sqrt{6} \, \mbox{arctanh} \frac{\tp}{2}
  \\
  \tilde{\B{g}}_{ab}  & = & ( 1-\frac{1}{4}\tp^2 ) \, \tg_{ab}
\end{eqnarray}
\end{subequations}
which is a bijective mapping from
$\tp \in ]-2,2[$ to $\tilde{\bar{\phi}}\in ]-\infty,\infty[$.
\\
Due to the following diagram, illustrating the above described
relations,  it is evident that there is a variable transformation
regularizing
the
unphysical equations for the Klein-Gordon field:
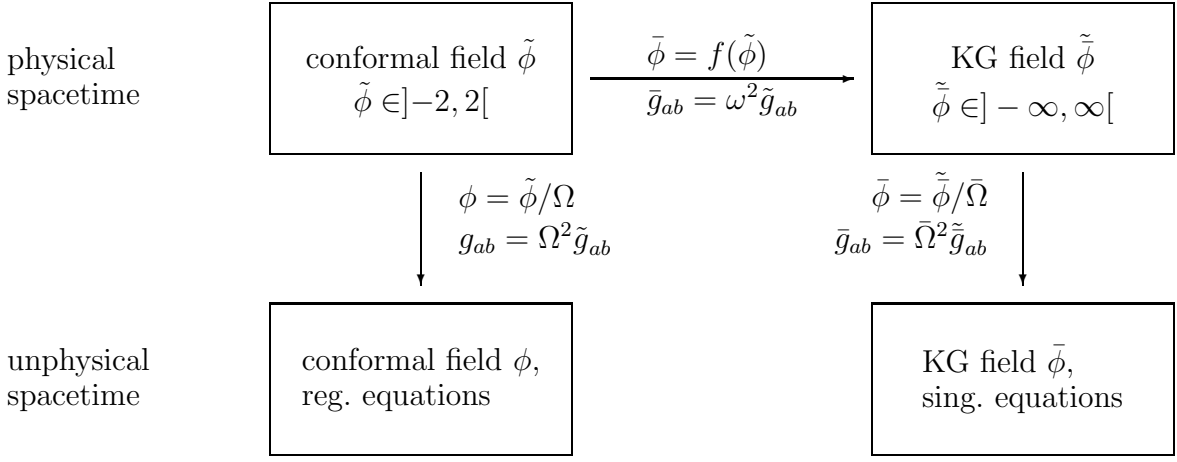
\begin{center}
\unitlength1cm
\begin{picture}(15,6.5)
\addcontentsline{lof}{figure}{{\string\numberline\space{}Konforme
    \protect"Aquivalenz von Skalarfeldern}}
  %
  \put(0,0){\makebox(2.5,2)[l]{\shortstack[l]{unphysical\\spacetime}}}
  \put(3.5,0){\framebox(4,2){%
              \shortstack[l]{conformal field $\phi$,\\reg.\ equations}}}
  \put(11.5,0){\framebox(4,2){%
              \shortstack[l]{KG field $\bar{\phi}$,\\sing.\ equations}}}
  %
  \put(5.5,3.75){\vector(0,-1){1.5}}
    \put(6,2.75){\shortstack[l]{$\phi=\tp/\Omega$\\$g_{ab}=\Omega^2\tg_{ab}$}}
  \put(13.5,3.75){\vector(0,-1){1.5}}
    \put(11,2.75){\shortstack[r]{$\bar{\phi}=\tilde{\bar{\phi}}/\bar{\Omega}$\\
                        $\bar{g}_{ab}=\bar{\Omega}^2\tilde{\bar{g}}_{ab}$}}
  %
  \put(0,4){\makebox(3,2)[l]{\shortstack[l]{physical\\spacetime}}}
  \put(3.5,4){\framebox(4,2){\shortstack{conformal field $\tp$\\
    $\tp \in ]{-2},{2}[$}}}
  \put(11.5,4){\framebox(4,2){\shortstack{KG field
        $\tilde{\bar{\phi}}$\\ $\tilde{\bar{\phi}}\in
        ]-\infty,\infty[$}}}
  %
  \put(7.75,5){\vector(1,0){3.5}}
  \put(8.5,5.2){\mbox{$\bar\phi=f(\tp)$}}
  \put(8.5,4.6){\mbox{$\bar g_{ab}=\omega^2\tg_{ab}$}}
\end{picture}
\end{center}
By mapping an arbitrary scalar field $\tilde{\bar{\bar \phi}}$ with
action (\ref{ScalarAction}) to the Klein-Gordon field $\tilde{\bar
  \phi}$ and then to the conformally invariant scalar field $\tilde
\phi $ regular equations for $\bar{\bar \phi}$ are obtained.
\subsection{The hyperboloidal initial value problem}
Since $A(\phi)>0$ on $\bar S$ all scalar field models connected
by transformation (\ref{Transf}) to a
hyperboloidal initial value problem with a conformally invariant
scalar field as matter source are weakly asymptotically flat.
\\
For a massless KG model there is a one parameter gauge freedom in the
scalar field. If $\tilde{\bar{\phi}}$ is a solution, then so is
$\tilde{\bar{\phi}}+\tilde{\bar{\phi}}_0$ with
$\tilde{\bar{\phi}}_0=\const{}$, as the energy momentum tensor depends on
derivatives of $\tilde{\bar{\phi}}$ only. This can also be seen by
mapping a Klein-Gordon model to a Klein-Gordon model with
$\tilde{\bar{\phi}}_0\ne{}0$ and (\ref{Transf}). The analogue holds for
every considered scalar field model.
For the hyperboloidal initial value problem $\tp = \Omega
\phi{}$, therefore $\tp{}$ vanishes at \Scri{}, fixing the gauge in
$\tp{}$.
\\
In definition (\ref{HypInitValProbl}) $1-\frac{1}{4} \Omega^2
\phi^2 \mid_{\bar S} >0$ was assumed. But with the Bekenstein black
hole~\cite{Be75bh} a weakly asymptotically flat
solution is known where $A(\tp)$ vanishes on a regular
part of the spacetime. In this case the transformation gives a
possible extension of a massless Klein-Gordon scalar field solution beyond a
singularity -- the Klein-Gordon field $\widetilde{\bar\phi}$ and the
metric $\widetilde{\bar g}_{ab}$ degenerate there.
\vspace{0.5cm}
It is a pleasure for me to thank Helmut Friedrich, Bernd Schmidt, and
J\"urgen Ehlers for the very helpful discussions during the grow of
this work which is part of my Ph.~D.\ thesis.

\begin{appendix}
\section{Notation}
The signature of the Lorentzian metric $g_{ab}$ is $(-,+,+,+)$.
\\
Whenever possible I use abstract indices as described
in~\cite[chapter~2]{PeR84SA}. Small Latin letters denote abstract
indices, underlined
small Latin letters are frame indices. For the components of a tensor with
respect to coordinates small Greek letters are used. The frame
$\left(\frac{\partial}{\partial x^\mu}\right)^a$ is constructed from
the coordinates $x^\mu $, $e_{\f{i}}{}^a$ denotes an arbitrary frame.
In this notation $v_a$ is a covector, $v_{\f{i}}$ a scalar, namely
$v_a\, e_{\f{i}}{}^a$.
\\
$v(f)$ is defined to be the action of the vector $v^a$ on the function
$f$, i.e.\ for every covariant derivative $\nabla_a$: $t(f)=t^a\,\nabla_a f$.
\\
The transformation between abstract, coordinate, and frame indices is
done by contracting with $e_{\f{i}}{}^a$ and $e_{\f{i}}{}^\mu $. All
indices may be raised and lowered with the metric $g_{AB}$ and the
inverse $g^{AB}$. $g^{AC}\,g_{CB} = \delta^A{}_B$, $A$ and $B$ are
arbitrary indices, e.g.\  $e_{\f{i}a}=g_{ab}\,e_{\f{i}}{}^b$ and
$e^{\f{i}}{}_a=g^{\f{i}\f{j}}\,e_{\f{i}a}$.
\\
For a frame $e_{\f{i}}{}^a$ and a covariant derivative
$\nabla_a$ the Ricci rotation coefficients
are defined as
\begin{equation*}
  \gamma^a{}_{\f{i}\f{j}} := e_{\f{i}}{}^b \nabla_b e_{\f{j}}{}^a.
\end{equation*}
{}From this definition follows
\begin{equation*}
  e_{\f{i}}{}^a \, e^{\f{j}}{}_b \, (\nabla_a t^b) =
  e_{\f{i}}(t^{\f{j}}) + \gamma^{\f{j}}{}_{\f{i}\f{k}} \, t^{\f{k}}.
\end{equation*}
\\
With respect to a coordinate frame $e_{\mu}{}^a \equiv
\left(\frac{\partial}{\partial x^\mu}\right)^a$  the components
$\gamma^\lambda{}_{\mu\nu}$ are the Christoffel symbols
$\Gamma^\lambda{}_{\mu\nu}$.
\\
The torsion $T^a{}_{bc}$ is defined by
\begin{equation*}
  \nabla_a\nabla_b f - \nabla_b\nabla_a f = - T^c{}_{ab} \, \nabla_c f,
\end{equation*}
the Riemann tensor $R_{abc}{}^d$ by
\begin{equation*}
  \nabla_a\nabla_b\omega_c - \nabla_b\nabla_a\omega_c =
  R_{abc}{}^d \, \omega_d  - T^{d}{}_{ab} \, \nabla_d \omega_c.
\end{equation*}
Contraction gives the Ricci tensor,
\begin{equation*}
  R_{ab} = R_{acb}{}^c,
\end{equation*}
and the Ricci scalar
\begin{equation*}
  R = R_{ab}\, g^{ab}.
\end{equation*}
The Einstein tensor is given by
\begin{equation*}
  G_{ab} = R_{ab} - \frac{1}{2} \, R \, g_{ab}.
\end{equation*}
\\
The speed of light $c$ is set to $1$ as the gravitational constant
$\kappa $ in $G_{ab}=\kappa\, T_{ab}$.
\end{appendix}

\end{document}